\documentclass[a4paper,12pt]{article}
\usepackage{graphicx,amssymb,bm,latexsym}
\newcommand{\bea}{\begin{eqnarray}}
\newcommand{\ena}{\end{eqnarray}}
\newcommand{\vs}[1]{\vspace{#1 mm}}
\newcommand{\hs}[1]{\hspace{#1 mm}}

\renewcommand{\c}{\gamma}
\renewcommand{\d}{\delta}

\newcommand{\dsl}{\pa \kern-0.5em /}

\newcommand{\shalf}{\frac{1}{2}}
\newcommand{\pa}{\partial}

\newcommand{\nn}{\nonumber\\}
\newcommand{\p}[1]{(\ref{#1})}

\begin{document}

\begin{titlepage}

\begin{flushright}
OU-HET 542 \\
YITP-05-57
\end{flushright}

\vs{10}
\begin{center}
{\Large\bf Time-dependent Solutions with Null Killing Spinor
in M-theory and Superstrings}
\vs{15}

{\large
Takayuki Ishino,$^{a,}$\footnote{e-mail address: ishino@het.phys.sci.osaka-u.ac.jp}
Hideo Kodama$^{b,}$\footnote{e-mail address: kodama@yukawa.kyoto-u.ac.jp}
and
Nobuyoshi Ohta$^{a,}$\footnote{e-mail address: ohta@phys.sci.osaka-u.ac.jp}} \\
\vs{10}
$^a${\em Department of Physics, Osaka University,
Toyonaka, Osaka 560-0043, Japan}

$^b${\em Yukawa Institute for Theoretical Physics, Kyoto University,
Kyoto 606-8502, Japan}

\vs{15}
{\bf Abstract}
\end{center}
\vs{5}

Imposing the condition that there should be a null Killing spinor with all
the metrics and background field strengths being functions of the light-cone
coordinates, we find general 1/2 BPS solutions in $D=11$ supergravity,
and discuss several examples. In particular we show that the linear
dilaton background is the most general supersymmetric solution without
background under the additional requirement of flatness in the string frame.
We also give the most general solutions for flat spacetime in the string frame
with RR or NS-NS backgrounds, and they are characterized by a single function.

\end{titlepage}
\newpage
\renewcommand{\thefootnote}{\arabic{footnote}}
\setcounter{page}{2}


Study of time-dependent solutions in string theories is an important
subject for its application to cosmology and understanding our
spacetime~\cite{time1,time2}. Among these, the solutions with partial supersymmetry
(BPS solutions) are important because these allow us to discuss
nonperturbative regions of our spacetime. It is known that
the requirement of unbroken partial supersymmetry restricts the solutions to
those with null or time-like Killing spinors~\cite{GP}. In this paper,
we focus on those solutions with null Killing spinors.

Recently such time-dependent solutions have been studied in the linear
dilaton background in the null direction with $\frac12$
supersymmetry~\cite{bb1,bb2}. (Related solutions of null branes are
considered in Ref.~\cite{null}.)
This is interesting not only in its BPS
property but also in the possibility of realization in Matrix theory,
which allows nonperturbative study of the solution. It has been
suggested that the problem of singularity
in the spacetime is under control in this setting.

The solution studied in \cite{bb1} does not involve any background in the forms
which are present in the string theories. This is certainly a simplification
but the question then arises if the results are restricted to only this
very special solution or they are valid for other related solutions
with possible backgrounds.
To answer this question, we have to know how a large class of solutions
of this type are allowed in string theories. The purpose of this paper
is to give more general class of solutions with nontrivial backgrounds.


Let us start with the general action for gravity coupled to a dilaton
$\phi$ and an $n$-form field strength:
\bea
I = \frac{1}{16 \pi G_D} \int d^D x \sqrt{-g} \left[
 R - \shalf (\pa \phi)^2 - \frac{1}{2 n!} e^{a \phi} F_{n}^2 \right].
\label{act}
\ena
This action describes the bosonic part of $D=11$ or $D=10$ supergravities;
we simply drop $\phi$ and put $a=0$ and $n=4$ for $D=11$, whereas we
set $a=-1$ for the NS-NS 3-form and $a=\shalf(5-n)$ for the form coming
from the R-R sector. To describe more general supergravities in lower
dimensions, we should include several scalars and forms,
but for simplicity we disregard this complication in this paper.

We are interested in the metric
\bea
ds^2 = -2 e^{2u_0} du dv + \sum_{i=2}^{D-1} e^{2u_i} (dx^i)^2,
\label{met}
\ena
where all metrics are assumed to be functions of $u=(t-x^1)/\sqrt{2}$ and
$v=(t+x^1)/\sqrt{2}$. Though we could consider more general metric in
the $(u,v)$ part, we can always put the metric in the above form without loss
of generality since the two-dimensional metric is (locally) conformally flat.
The components of the Ricci tensor are
\bea
R_{uu} &=& - \sum_{i=2}^{D-1} \pa_u^2 u_i + 2 \pa_u u_0 \sum_{i=2}^{D-1}\pa_u u_i
 - \sum_{i=2}^{D-1} (\pa_u u_i)^2, \nn
R_{uv} &=& - \pa_u \pa_v \Big(2u_0+\sum_{i=2}^{D-1} u_i\Big)
- \sum_{i=2}^{D-1} \pa_u u_i \pa_v u_i, \nn
R_{vv} &=& - \sum_{i=2}^{D-1} \pa_v^2 u_i +2 \pa_v u_0 \sum_{i=2}^{D-1}\pa_v u_i
 -\sum_{i=2}^{D-1} (\pa_v u_i)^2, \nn
R_{ij} &=& \d_{ij} e^{2(u_i-u_0)} \Big[
2 \pa_u \pa_v u_i + \pa_v u_i \sum_{k=2}^{D-1}\pa_u u_k
+ \pa_u u_i \sum_{k=2}^{D-1}\pa_v u_k \Big].
\ena
We consider mainly the $D=11$ supergravity and so there is only an $n=4$ form.
We take the background
\bea
F = (\pa_u E du + \pa_v E dv)\wedge dx^2 \wedge dx^3 \wedge dx^4,
\ena
leading to
\bea
F^2 = - 4! \cdot 2 (\pa_u E \pa_v E) e^{-2(u_0 +U)},
\ena
where $U=u_2+u_3+u_4$. This is an electric background and we could also
consider magnetic background, but that is basically the same as the electric
case with the replacement
\bea
g_{\mu\nu} \to g_{\mu\nu}, \quad
F_{n} \to e^{a \phi}*\! F_{n}, \quad
\phi \to - \phi.
\label{sdual}
\ena
This is due to the S-duality symmetry of the original system~\p{act}.
So we do not have to consider it separately.

The field equations following from the general action~\p{act} are as follows:
\bea
&&\hs{-4} \label{fs}
0 = \pa_{\mu_1} (\sqrt{-g} e^{a \phi} F^{\mu_1 2 3 4}) \nn
&& = -\pa_u \left\{ e^{a \phi-U+\sum_{i=5}^{D-1} u_i} \pa_v E \right\}
-\pa_v \left( e^{a \phi-U+\sum_{i=5}^{D-1} u_i} \pa_u E \right),\\
&&\hs{-10}
\frac{1}{\sqrt{-g}} \pa_u\Big[ e^{\sum_{i=2}^{D-1} u_i} \pa_v \phi \Big]
+\frac{1}{\sqrt{-g}} \pa_v\Big[ e^{\sum_{i=2}^{D-1} u_i} \pa_u \phi \Big]
= a e^{a \phi -2u_0-2U} \pa_v E \pa_u E, \\
&&\hs{-10} \label{uu}
-\sum_{i=2}^{D-1}\pa_u^2 u_i +2\pa_u u_0 \sum_{i=2}^{D-1}\pa_u u_i
- \sum_{i=2}^{D-1}(\pa_u u_i)^2
= \frac12 (\pa_u \phi)^2 +\frac12 e^{a \phi-2 U}(\pa_u E)^2, \\
&& \hs{-10} \label{uv}
-\pa_u \pa_v(2u_0 +\sum_{i=2}^{D-1} u_i)
- \sum_{i=2}^{D-1} \pa_u u_i \pa_v u_i
= \frac12 \pa_u \phi \pa_v \phi + \frac{D-8}{2(D-2)} e^{a \phi-2U}
 \pa_u E \pa_v E, \\
&& \hs{-10}
-\sum_{i=2}^{D-1} \pa_v^2 u_i +2\pa_v u_0 \sum_{i=2}^{D-1}\pa_v u_i
- \sum_{i=2}^{D-1} (\pa_v u_i)^2
= \frac12 (\pa_v \phi)^2 + \frac12 e^{a \phi -2U} (\pa_v E)^2, \label{vv} \\
&& \hs{-10}
e^{2(u_i -u_0)} \Big[ 2\pa_u \pa_v u_i +\pa_v u_i \sum_{k=2}^{D-1}\pa_u u_k
+\pa_u u_i \sum_{k=2}^{D-1}\pa_v u_k \Big]
= -\frac{D-5}{D-2} e^{a \phi+ 2(-u_0+u_i-U)} \pa_u E \pa_v E, \nn
&& \hs{30}i=2,3,4, \\
&& \hs{-10}
e^{2(u_i -u_0)} \Big[ 2\pa_u \pa_v u_i +\pa_v u_i \sum_{k=2}^{D-1}\pa_u u_k
+\pa_u u_i \sum_{i=2}^{D-1} \pa_v u_k \Big]
= \frac{3}{D-2} e^{a \phi+ 2(-u_0+u_i-U)} \pa_u E \pa_v E, \nn
&& \hs{30} i=5, \cdots,D-1.
\label{ein}
\ena

{}From now on, we restrict our discussions to $D=11$ supergravity
for simplicity, but it is straightforward to repeat our analysis for
other theories given our field equations.
The supersymmetry transformation for $D=11$ supergravity without dilaton is
\bea
\delta\psi_\mu=\Big[\partial_\mu+\frac{1}{4}\omega^{ab}_\mu\Gamma_{ab}
+\frac{1}{288}({\Gamma_\mu}^{\nu\rho\sigma\tau}-8{\delta_\mu}^\nu
\Gamma^{\rho\sigma\tau})
F_{\nu\rho\sigma\tau}\Big]\zeta,
\ena
where $\Gamma_{ab}$ are antisymmetrized gamma matrices and
\bea
\omega_{\mu ab}\equiv\frac{1}{2}e^\nu_a(\pa_\mu e_{b\nu}-\pa_\nu
e_{b\mu})-\frac{1}{2}e^\nu_b(\partial_\mu e_{a\nu}-\partial_\nu e_{a\mu})
-\frac{1}{2}e^\rho_ae^\sigma_be^c_\mu(\partial_\rho e_{c\sigma}
-\partial_\sigma e_{c\rho}),
\ena
is the spin connection. The Killing spinor equations are obtained
by setting these to zero:
\bea
\d \psi_u &=& \Big[ \pa_u -\frac16 \c_{234}e^{-U} \pa_u E
+\c_{-+} \Big( \frac12 \pa_u u_0+\frac1{12} \c_{234} e^{-U} \pa_u E \Big)
\Big] \zeta,
\label{u} \\
\d \psi_v &=& \Big[ \pa_v -\frac16 \c_{234}e^{-U} \pa_v E
-\c_{-+} \Big( \frac12 \pa_v u_0
+\frac{1}{12} \c_{234} e^{-U} \pa_v E \Big)\Big]\zeta,
\label{v} \\
\d \psi_i &=& \Big[ \pa_i +\c_{-i} e^{u_i-u_0} \Big(\frac12 \pa_u u_i
-\frac1{6}  \c_{234}e^{-U} \pa_u E \Big) \nn
&& \qquad +\c_{+i} e^{u_i-u_0} \Big(\frac12 \pa_v u_i
-\frac1{6} \c_{234} e^{-U} \pa_v E \Big) \Big]\zeta, \quad i=2, 3, 4,
\label{a} \\
\d \psi_i &=& \Big[ \pa_i +\c_{-i} e^{u_i-u_0} \Big(\frac12 \pa_u u_i
+\frac1{12} \c_{234}e^{-U} \pa_u E \Big) \nn
&& \qquad +\c_{+i} e^{u_i-u_0} \Big(\frac12 \pa_v u_i
+\frac1{12} \c_{234} e^{-U} \pa_v E \Big) \Big]\zeta, \quad i=5, \ldots, 10,
\label{b}
\ena
where $\c_{-+}=\frac{1}{2}(\c_- \c_+ -\c_+ \c_-)$ and
$\c_\pm =\frac{1}{\sqrt{2}}(\c_0 \pm \c_1)$.

We look for supersymmetric solutions for which eqs.~\p{u}--\p{b} all vanish.
We assume that $\zeta$ can depend only on $u$ and $v$.
It then follows from eq.~\p{b} that
\bea
[\gamma_- (6\partial_u u_i + \gamma_{234}e^{-U} \partial_u E)+
\gamma_+(6\partial_v u_i + \gamma_{234}e^{-U}\partial_v E)]\zeta=0.
\ena
Multiplying $\gamma_-$ or $\gamma_+$, we find that nontrivial solutions can
be obtained only if either $\gamma_-\zeta=0$ or $\gamma_+\zeta=0$, unless
all $u_i$ and E are constant. We choose
\bea
\c_- \zeta=0.
\label{bps1}
\ena
Then we have
\bea
\partial_v E=\pa_v u_i=0,\quad
(i=2, \ldots,10),
\label{bps2}
\ena
where the last condition for $i=2, 3, 4$ follows from eq.~\p{a}.
The $uv$ component of the Einstein eq.~\p{uv} then requires that
\bea
\pa_u \pa_v u_0 =0.
\ena
This means that $u_0=P(u)+Q(v)$. We could then set $u_0=0$ by a gauge choice,
but we find it more instructive to keep the $u$-dependence;
we take $u_0=u_0 (u)$ in what follows.

Note that eq.~\p{bps1} means that $\c_{-+}\zeta=-\zeta$.
Eqs.~\p{u} and \p{v} now reduce to
\bea
\d \psi_u &=& \Big[ \pa_u -\frac14 \c_{234}e^{-U} \pa_u E
- \frac12 \pa_u u_0 \Big] \zeta =0,\nn
\d \psi_v &=& \pa_v \zeta=0.
\label{v1}
\ena
Let us define
\bea
\zeta = \exp\Big[\c_{234} h  + \frac{1}{2}u_0(u)  \Big]
\zeta_0,
\ena
where
\bea
\pa_u h = \frac{1}{4} e^{-U} \pa_u E, \quad
\pa_v h = 0.
\ena
Note that the Majorana spinor condition $\zeta= C (\bar \zeta)^T$ is
satisfied if $\zeta_0$ is a Majorana spinor.
Namely for $\zeta =e^{\c_{234}h} \zeta_1$ with a Majorana
spinor $\zeta_1$,
we have $C\bar \zeta^T = C(\zeta_1^\dagger e^{\c_{432}h} i\c_0)^T
=C i\c_0^T e^{\c_2^T \c_3^T \c_4^T h} \zeta_1^*
=- i\c_0 e^{- \c_2 \c_3 \c_4 h} C\zeta_1^*
=- e^{\c_{234} h} i\c_0 C \zeta_1^*
= e^{\c_{234} h}\zeta_1 =\zeta$ where we have used the Majorana property of
$\zeta_1$: $\zeta_1= C i\c_0^T \zeta_1^* =-i\c_0 C \zeta_1^*$.
We also note that the integrability condition $(\pa_u\pa_v-\pa_v\pa_u) h=0$
is satisfied due to \p{bps2}. Then, eq.~\p{v1} is equivalent to
\bea
\pa_u \zeta_0 = \pa_v \zeta_0=0.
\label{bps3}
\ena
Hence, eq.~\p{bps1} is the only relevant condition on the Killing spinor and
the number of remaining supersymmetry is $\frac12$.

The $uu$ component of the Einstein eq.~\p{uu} reduces to\footnote{
The dilaton is absent for $D=11$ supergravity.}
\bea
\sum_{i=2}^{10}\pa_u^2 u_i + \sum_{i=2}^{10} (\pa_u u_i)^2
+ \frac{1}{2} e^{-2U}(\pa_u E)^2 =2\pa_u u_0 \sum_{i=2}^{10}\pa_u u_i,
\label{bps4}
\ena
and others are trivial after using the conditions~\p{bps2}.
Because these are all functions of $u$ only, this equation is an ordinary
differential equation for 11 functions $u_i \; (i=0,2,\cdots,10)$ and $E$.
We can regard eq.~\p{bps4} as determining $E$ for given metrics.
This in fact gives a very large class of solutions in $D=11$
supergravity, generalizing those given in Refs.~\cite{bb1,bb2}.

Given a solution to eq.~\p{bps4}, the metric for our spacetime is given
in eq.~\p{met}. Upon dimensional reduction to 10 dimensions, we get
the string frame metric by
\bea
ds_{11}^2 = e^{-2\phi/3} ds_{st}^2 + e^{4\phi/3} (dx^{10})^2,
\ena
where
\bea
ds_{st}^2 = e^{u_{10}} \Big[-2 e^{2u_0} du dv +\sum_{i=2}^9 e^{2u_i} (dx^i)^2\Big],
\quad
\phi=\frac{3}{2} u_{10}.
\label{string}
\ena
The Einstein frame metric is then obtained as
\bea
ds_E^2 &=& e^{-\phi/2} ds_{st}^2 \nn
&=& e^{u_{10}/4} \Big[-2 e^{2u_0} du dv +\sum_{i=2}^9 e^{2u_i} (dx^i)^2\Big].
\ena

This is our main result. The solutions we have obtained appear slightly
different from those in Ref.~\cite{GP} though ours are more explicit.
It can be checked that the apparent difference is due to the different basis.
We have confirmed that they are consistent if we make a Lorentz transformation.

Now in order for the theory to be solvable, the metric in the string frame
should be of that type. The simplest case is when it is flat.

Let us now discuss some interesting cases.

{\bf Case (i): Flat metric in the string frame.}\\
To have a flat metric in the string frame, we find from eq.~\p{string}
\bea
u_0=u_i=-\frac{u_{10}}{2} \equiv f.
\label{flat}
\ena
Substituting this into eq.~\p{bps4}, we get
\bea
f'' =-\frac{1}{12}e^{-6f}(E')^2.
\label{f1}
\ena
If we do not introduce the background, this yields
\bea
f= c u,
\label{lind}
\ena
where we have dropped one integration constant which can be absorbed into
the rescaling of coordinates. This is the linear dilaton, and most general
solution which gives the flat space in the string frame without any background.
Note that this is by no means trivial theory in the Einstein
frame as long as the dilaton is nontrivial.

If we introduce background, any function $f$ satisfying \p{f1} is allowed.
For instance, the next simplest case is a constant field strength:
\bea
E'= 6a.
\ena
We get from eq.~\p{f1}
\bea
e^{3f}= \frac{a}{2c} \left( \frac{1}{b} e^{3cu} - b e^{-3cu} \right).
\ena
Setting $b=a/(2c)$, we reproduce the solution~\p{lind} in the $a\to 0$ limit.
If we put $b=\pm 1$ and take the $c\to 0$ limit, we get another simple solution
\bea
f = \frac13 \ln(3 a |u|).
\ena
Study of string theory in this background may not be difficult because it has
flat metric in the string frame.

{\bf Case (ii): Linear dilaton in the light-like direction.}\\
As a generalization of the simple linear dilaton background obtained above,
we can consider
\bea
\phi=-Qu, \quad
u_{10}=-\frac23 Qu.
\label{f2}
\ena
If we also assume $u_0$ and $u_i$ are all linear in $u$,
\bea
u_i = c_i u, \quad
(i=0, 2, \cdots 9),
\ena
\p{bps4} gives
\bea
\sum_{i=2}^{9} c_i^2 +\frac{4}{9}Q^2 - 2 c_0 \Big(\sum_{i=2}^9 c_i -\frac23 Q \Big)
=-\frac12 (E')^2 e^{-2 (c_2+c_3+c_4)u} .
\label{lin}
\ena
If we do not introduce the background and take all $c_i$ equal,
\p{lin} leads to
\bea
c_i=\frac13 Q,~ -\frac16 Q.
\ena
The first one coincides with the linear dilaton in the flat metric in the
string frame obtained above for $c=\frac13 Q$, and is the linear dilaton
background considered in Ref.~\cite{bb1}.
However, the metric in the string frame is not trivial.
The second one, also considered in Ref.~\cite{bb2}, has nontrivial
background in the string frame, but we might study the string theory
in this background by Matrix theory counterpart.
For general $c$, we still have a solution with background
\bea
E' = \pm \frac{2}{3}\sqrt{2(3c-Q)(6c+Q)} \; e^{3cu}.
\ena

Here we comment on the general but formal solutions of \p{f1}.
Since both $E$ and $f$ are functions of $u$, we could consider $E$ as
an implicit function of $f$. The general solution is then given as
\bea
u=\int df \exp\Bigg[\frac{1}{12} \int df e^{-6f} \left(\frac{ dE}{df} \right)^2
\Bigg].
\ena

{\bf Case (iii): Solutions with 2 functions.}\\
If we set
\bea
u_0 = u_2 = \cdots = u_{10-d}=f, ~~
u_{11-d} = \cdots = u_{10}=g,
\ena
eq.~\p{bps4} gives
\bea
(9-d)(f'^2 -f'')+d(2f' g' -g'^2 -g'')=\frac{1}{2} e^{-6f}(E')^2,
\ena
where we have assumed $d \leq 6$. For $g=E'=0$, we get $f''-f'^2=0$ and
$f=-\ln u$ (the integration constants can be absorbed to the shift of $u$
and rescaling of the coordinates)~\cite{bb2}. After a change of variable,
this yields the nontrivial background similar to the orbifold discussed
in Ref.~\cite{time1}.

\def\cases#1{\left\{ \begin{array}{ll}#1\end{array}\right.}

If we do not consider orbifold, this solution $f=-\ln u$ corresponds to flat
space. We can have more general solutions of this type
interpolating two asymptotically flat spaces. If we set
\begin{equation}
u_0=0,\ u_2=u_3=u_4=f,\ u_5=\cdots=u_{10-d}=g,\
u_{11-d}=\cdots=u_{10}=0,
\end{equation}
eq.~\p{bps4} gives
\begin{equation}
3e^{-f}(e^f)'' + (6-d) e^{-g}(e^g)''=-\frac{1}{2}e^{-6f} (E')^2.
\end{equation}
Note that for $d\ge1$, the metrics in the string frame and in the Einstein frame
are identical.

In particular, for $f=g$, this reduces to
\begin{equation}
e^{-f}(e^f)'' =-\frac{1}{2(9-d)} e^{-6f} (E')^2.
\end{equation}
Because a spacetime with $e^f,e^g \propto u$ or constant is (locally) flat
in the gauge $u_0=0$, any solution to this equation yields a spacetime that
is asymptotically flat at $u \rightarrow \pm \infty$, provided $E'$ does not
vanish except in a finite range with respect to $u$ or falls off sufficiently
rapidly as $u \rightarrow \pm\infty$. Such solutions can be regarded as
representing a decompactification transition, so they may deserve further
investigations.

For example, the solution
\begin{equation}
e^f= 1+u -(1+u^4)^{1/4},\ (u>0),
\end{equation}
is obtained for
\begin{equation}
E'=\frac{\sqrt{6(9-d)}u}{(1+u^4)^{7/8}} \{1+u-(1+u^4)^{1/4}\}^{5/2},
\end{equation}
which behaves as
\begin{equation}
E' \propto \cases{ u^{-5/2} & \mbox{for } u \to \infty,\\
                   u^{7/2} & \mbox{for } u \to 0.}
\end{equation}
The Riemann curvature components with respect to the normalized null basis are
proportional to
\begin{equation}
f''+(f')^2 = -\frac{3u^2}{(1+u^4)^{7/4}[1+u-(1+u^4)^{1/4}]},
\end{equation}
which behaves for $u\rightarrow\infty$ and $u\sim 0$ as
\begin{equation}
f''+(f')^2 \simeq \cases{ -3 u^{-5} & \mbox{for } u\to \infty,\\
                          -3u & \mbox{for } u\to 0.  }
\end{equation}
Hence, the solution is flat for $u\to \infty$ and $u=0$, and as a
consequence it can be smoothly extended to the completely flat region in
$u\le0$.

{\bf Case (iv): Solutions with NS-NS backgrounds.}\\
The solutions we discussed up to this point except for those without backgrounds
involve RR backgrounds, which are not easy to deal with. If we compactify
in the 4-th direction, however, we can get solutions with NS-NS 2-form $B$:
\bea
ds_{st}^2 = e^{u_{4}} \Big[-2 e^{2u_0} du dv +\sum_{i=2(\neq 4)}^{10}
e^{2u_i} (dx^i)^2\Big],
\quad
\phi=\frac{3}{2} u_{4}, \quad
B=E dx^2 \wedge dx^3.
\label{stringns}
\ena
To have flat metric in the string frame, we find from eq.~\p{stringns}
\bea
u_0=u_i=-\frac{u_{4}}{2} \equiv f.
\label{flatns}
\ena
Substituting this into eq.~\p{bps4}, we get
\bea
f''= -\frac{1}{12}(E')^2.
\label{f1ns}
\ena
This yields further interesting class of solutions whose spectrum
can be investigated with string actions with these backgrounds.
In fact we could also get similar results for linear dilaton and
solutions with 2 functions easily.
In particular we can have constant NS-NS 2-form background together with
linear dilatons. It is known that such a constant NS-NS background yields
noncommutative theory~\cite{nonc}. It is extremely interesting to
examine what implications these backgrounds may have on the singular
behavior in the spacetime.


It would be also interesting to explore the possible application of
our solutions to cosmological models.
Whether our general solutions have
interesting applications to these problems remains to be seen.

In summary, imposing the condition that there should be a null Killing
spinor with all the metrics and background field strengths being functions
of the light-cone coordinates, we have found general 1/2 BPS solutions in $D=11$
supergravity, and discussed several examples. In particular we have shown
that the linear dilaton background is the most general supersymmetric
solution without background with the additional requirement of flatness
in the string frame. We have also given the most general solutions for
flat spacetime in the string frame with RR or NS-NS backgrounds, and they are
characterized by a single function $f$ or $E$ related as \p{f1} or \p{f1ns}.
The fact that the solutions are flat in the string frame suggests that
they could be put in the Matrix theory, which allows nonperturbative study.
Considering the difficulty in the study of string theories in
time-dependent backgrounds, it should be very interesting to study these
problems. We hope to report on these issues in the near future.

\section*{Acknowledgement}
The work of HK and NO was supported in part by the Grant-in-Aid for
Scientific Research Fund of the JSPS No. 15540267 and No. 16540250.

\newcommand{\NP}[1]{Nucl.\ Phys.\ B\ {\bf #1}}
\newcommand{\PL}[1]{Phys.\ Lett.\ B\ {\bf #1}}
\newcommand{\CQG}[1]{Class.\ Quant.\ Grav.\ {\bf #1}}
\newcommand{\CMP}[1]{Comm.\ Math.\ Phys.\ {\bf #1}}
\newcommand{\IJMP}[1]{Int.\ Jour.\ Mod.\ Phys.\ {\bf #1}}
\newcommand{\JHEP}[1]{JHEP\ {\bf #1}}
\newcommand{\PR}[1]{Phys.\ Rev.\ D\ {\bf #1}}
\newcommand{\PRL}[1]{Phys.\ Rev.\ Lett.\ {\bf #1}}
\newcommand{\PRE}[1]{Phys.\ Rep.\ {\bf #1}}
\newcommand{\PTP}[1]{Prog.\ Theor.\ Phys.\ {\bf #1}}
\newcommand{\PTPS}[1]{Prog.\ Theor.\ Phys.\ Suppl.\ {\bf #1}}
\newcommand{\MPL}[1]{Mod.\ Phys.\ Lett.\ {\bf #1}}
\newcommand{\JP}[1]{Jour.\ Phys.\ {\bf #1}}

\end{document}